\documentclass[cameraready]{Interspeech}

\usepackage{microtype}
\usepackage{subfigure}
\usepackage{multirow}
\usepackage{hyperref}
\usepackage{mathtools}
\usepackage[capitalize,noabbrev]{cleveref}
\usepackage[disable]{todonotes}
\usepackage{amssymb}
\usepackage{pifont}
\newcommand{\xmark}{\ding{55}}
\newcommand{\ceil}[1]{\left\lceil {#1} \right\rceil}

\title{Benchmarking Language Modeling for Lossless Compression of Full-Fidelity Audio}

\author[affiliation={1}, correspondingauthor, equalcontribution]{Phillip}{Long}
\author[affiliation={1}, equalcontribution]{Zachary}{Novack}
\author[affiliation={2}]{Chris}{Donahue}

\address{
    $^1$ University of California, San Diego, Computer Science and Engineering Department \\
    $^2$ Carnegie Mellon University, School of Computer Science
}

\email{}


\definecolor{zgreen}{RGB}{29, 185, 84}

\usepackage{comment}

\begin{document}

\maketitle

\begin{abstract}
    Autoregressive ``language'' models (LMs) trained on raw waveforms can be repurposed for \emph{lossless} audio compression, but prior work is limited to 8-bit audio, leaving open whether such approaches work for practical settings (16/24-bit) and can compete with existing codecs. We benchmark LM-based compression on full-fidelity audio across diverse domains (music, speech, bioacoustics), sampling rates (16kHz-48kHz), and bit depths (8, 16, 24-bit). Standard sample-level tokenization becomes intractable at higher bit depths due to vocabulary size (65K for 16-bit; 16.7M for 24-bit). We propose \emph{Trilobyte}, a byte-level tokenization schema for full resolution audio, improving vocabulary scaling from $O(2^{b})$ to $O(1)$ and enabling the first tractable 24-bit LM-based lossless compression. While LMs consistently outperform FLAC and yield state-of-the-art compression at 8-bit and 16-bit, we observe that compression gains become more modest as bit depth increases beyond 8-bit.
\end{abstract}

\section{Introduction}\label{sec:intro}

Recent work has demonstrated that neural-based approaches can yield dramatic improvements in \emph{lossy} audio compression \cite{zeghidour2021soundstream, kumar2023high, defossez2022high}, with learned codecs achieving an order of magnitude better compression rates than traditional codecs like MP3 \cite{brandenburg1999mp3} while maintaining comparable perceptual quality.
However, the potential of ML for \emph{lossless} audio compression remains largely unexplored at practical fidelities.
Autoregressive (AR) language models (LMs) offer a potential solution, as such models trained on raw audio samples can be repurposed as lossless compressors via arithmetic coding \cite{deletang2023language}, 
where compression rate improves as the model's log likelihood increases.
However, prior work has been constrained to 8-bit quantization at 16kHz sampling rates \cite{van2016wavenet,goel2022s,deletang2023language,li2025lossless, heurtel2024compression}.
This poses a practical problem as 8-bit, low sampling-rate audio is of limited downstream relevance---the perceptual quality is poor enough that audio is almost never distributed at this fidelity. 
Professional recording and production workflows universally operate 
on ``CD-quality'' audio (44.1kHz, 16-bit) or better (higher sample rates, 24-bit). 
Additionally, directly modeling higher bit-depth audio creates exponentially larger vocabularies for LMs ($2^{16} = 65{,}536$ tokens for 16-bit; $2^{24} = 16{,}777{,}216$ for 24-bit), rendering standard AR approaches computationally intractable.
Whether language model compression can thus scale to these full-fidelity regimes remains an open question.

To counteract the exponential scaling of vocabulary size as bit depth increases, we introduce \textbf{Trilobyte}, a 
byte-level tokenization scheme that 
improves vocabulary scaling from exponential $O(2^{b})$ to 
constant $O(1)$ 
in bit depth. 
We demonstrate that not only does Trilobyte improve compression rates for 16-bit audio relative to sample-level modeling, but it also enables the first tractable language model compression of 24-bit audio.
The Trilobyte tokenization scheme 
is compatible with any AR modeling framework---here we explore standard decoder-only Transformers specifically. 

We 
systematically 
evaluate LM-based compression, including pre-trained LLMs and models trained from scratch on audio with and without Trilobyte encoding, on full-fidelity audio across diverse domains (music, speech, and bioacoustic signals), sampling rates (16-48kHz), and at 8-, 16-, and 24-bit quantization. 
Our results primarily highlight that \emph{bit depth}, rather than sampling rate or data domain, is the limiting factor in LM-based compression. 
Specifically, while LM-based approaches can outperform the industry-standard lossless codec FLAC~\cite{FLAC} at 8-bit (217\% average improvement, consistent with prior work), the performance gap narrows substantially at higher bit depths (18\% improvement at 16-bit).

Our contributions include: (1) Trilobyte, enabling tractable 24-bit modeling through hierarchical tokenization with linear vocabulary scaling; (2) the first comprehensive benchmarking of LM compression on full-fidelity audio (16/24-bit) across diverse domains; and (3) evidence characterizing the performance gap between learned and traditional compressors across bit depths.
We make the code for Trilobyte publicly available.\footnote{\texttt{\url{https://github.com/pnlong/trilobyte-experiments}}}

\begin{figure*}[t]
    \centering
    \includegraphics[width=1.0\textwidth]{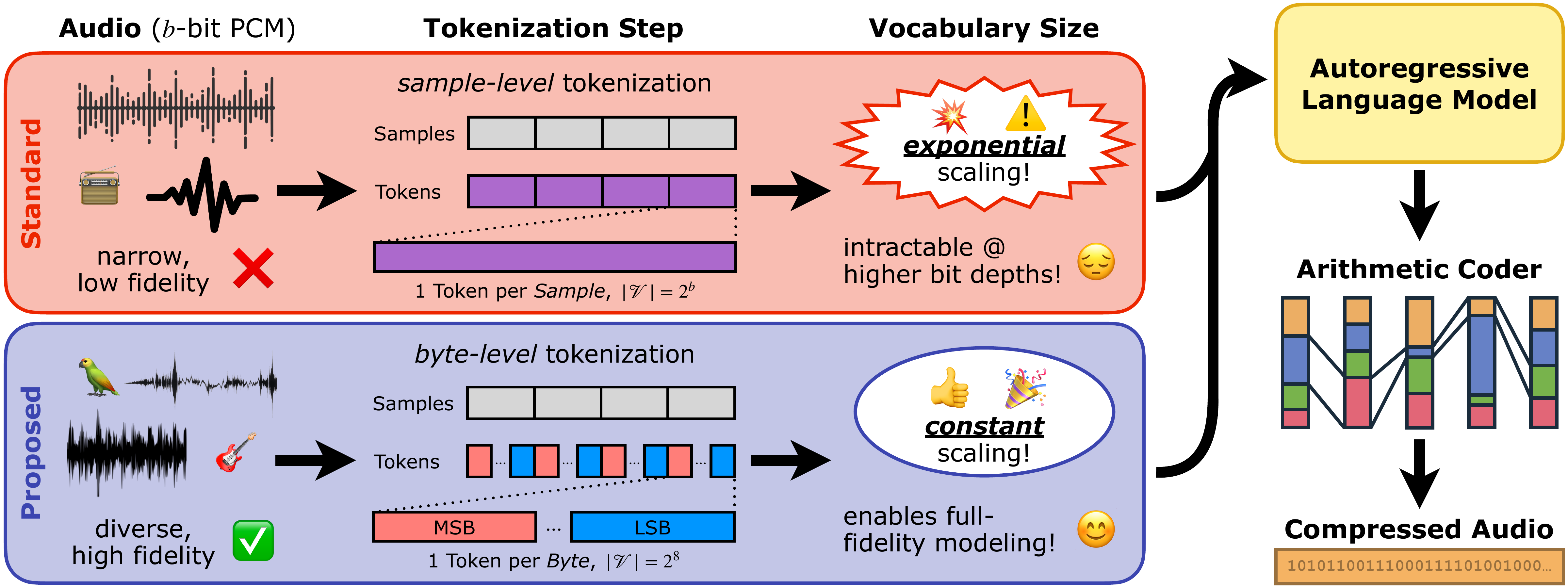}
    \caption{
        Tokenization strategies for language model compression.
        Standard sample-level tokenization (top) yields vocabulary size $|\mathcal{V}| = 2^{b}$. This exponential scaling inhibits modeling of industry-standard bit depths ($16$, $24$).
        Trilobyte's hierarchical byte-level tokenization (bottom) decomposes samples 
        into bytes, yielding constant $|\mathcal{V}| = 256$ regardless of bit depth (at the cost of increasing sequence length by $\ceil{b/8}$). 
        Both feed into an AR LM and arithmetic coder, but 
        Trilobyte 
        enables tractable 24-bit modeling.
    }
    \label{fig:diagram}
\end{figure*}

\section{Related Works}\label{sec:related_works}

\subsection{Traditional Lossless Audio Compression}

FLAC (Free Lossless Audio Codec) \cite{FLAC} is the de facto standard for lossless audio compression, achieving typical compression rates of 2x for CD-quality music \cite{vb2015losslesscomp}. 
In this work, we define the compression rate as $R = A / C(A)$, where $A$ and $C(A)$ represent the original and compressed sizes, respectively, and any $R > 1$ denotes a reduction in file size.
FLAC uses linear predictive coding to approximate audio in chunks and Rice coding \cite{rice1979some} to efficiently encode residuals \cite{van2024rfc}.
Although FLAC does attempt to exploit stereo redundancy via mid-side encoding, this yields only marginal compression rate improvements for CD-quality audio because the small chunk size of around 4,096 samples is insufficient to accommodate the stereo delays commonly present in music production, which prevents effective decorrelation of the left and right channels. 

\subsection{ML for Lossless Compression}

While recent work has shown that machine learning can dramatically improve \emph{lossy} audio compression \cite{zeghidour2021soundstream, kumar2023high, defossez2022high}, the lossless setting remains less explored.
Delétang et al. \cite{deletang2023language} and Li et al. \cite{li2025lossless} proposed using large language models like Llama \cite{touvron2023llama, touvron2023llama2} and Chinchilla \cite{hoffmann2022training} for general-purpose lossless compression via arithmetic coding \cite{pasco1976source, rissanen1976generalized}.
However, these works' exploration of \emph{audio} is limited to 8-bit Librispeech \cite{panayotov2015librispeech} and LJSpeech \cite{ljspeech}, where they outperform FLAC.
This 8-bit regime is not representative of real-world lossless audio applications, which typically require at least CD-quality (44.1kHz, 16-bit) formats.
Heurtel-Depeiges et al. \cite{heurtel2024compression} further demonstrate that small pre-trained Transformers can achieve competitive compression rates to FLAC on 8-bit audio.
Critically, none of these works explore whether compression gains extrapolate to higher bit depths.
Whether ML-based compression remains competitive at full-fidelity audio is the central question we investigate.

Prior work on AR modeling of raw audio waveforms \cite{van2016wavenet, mehri2016samplernn, goel2022s} has focused on generation rather than compression, operating primarily on 8-bit audio at 16--24kHz.
Despite this early interest in waveform-level modeling, such approaches have largely fallen out of favor for generation tasks, supplanted by methods that train on compressed audio tokens \cite{van2017neural, dhariwal2020jukebox, lakhotia2021generative}; however, raw waveform modeling remains promising for lossless compression.
Traditional approaches address the vocabulary explosion through $\mu$-law companding \cite{lewis1997law}, which reduces bit depth via non-linear quantization.
Recent work on byte-level modeling \cite{yu2023megabyte} explores tokenization strategies that could scale to larger vocabularies, though these have not been tested on higher bit-rate audio.
To our knowledge, no prior work has successfully trained AR models for 16- or 24-bit audio compression at CD-quality sample rates.

\section{Methods}\label{sec:methods}

AR models offer a fundamentally different paradigm from traditional codecs like FLAC: rather than using a bottleneck representation plus residuals, they directly model the probability distribution over the next sample given all previous samples.
The key insight is that any AR probabilistic model $P(x_{i} \mid x_{< i})$ over discrete sequences can be used with arithmetic coding \cite{pasco1976source, rissanen1976generalized, witten1987arithmetic} to achieve compression rates that approach the entropy of the data \cite{deletang2023language}.
Unlike FLAC's local chunk-based prediction, AR models can capture arbitrarily long-range dependencies in the audio signal, potentially discovering structure that linear prediction cannot.
The compression pipeline operates as follows: we train an AR model to predict $P(x_{i} \mid x_{< i})$ at each position, then during compression, we iteratively compute these probabilities and use arithmetic coding to encode the sequence into a bitstream. Decompression reverses this process, using the same AR model and arithmetic decoder to sequentially reconstruct each sample.

\subsection{Arithmetic Coding}

Unlike FLAC's use of Rice coding which assumes a fixed geometric distribution, arithmetic coding \cite{pasco1976source, rissanen1976generalized, witten1987arithmetic} can efficiently encode any sequence given an arbitrary probability distribution at each step.
The core principle is to encode an entire sequence into a single fractional number in $[0, 1)$ by iteratively narrowing an interval based on the cumulative probability distribution.
The encoding process begins with the interval $[0, 1)$; for each symbol $x_{i}$, we partition the current interval according to $P(x_{i} \mid x_{< i})$ and select the sub-interval corresponding to the observed symbol.
The final interval uniquely identifies the sequence, and we output a binary representation of any number within this interval.

Arithmetic coding achieves theoretical optimality, approaching the Shannon entropy \cite{shannon1948mathematical}.
The key advantage is the ability to exploit arbitrary probability distributions from the AR model, not just geometric ones.
This creates a direct connection to language models: the cross-entropy loss (negative log-likelihood) minimized during training directly corresponds to the expected coding length \cite{deletang2023language, heurtel2024compression}.
Specifically, average per-token log likelihood ${b_{\theta} = -\frac{1}{N} \Sigma_{i=1}^{N} \log_2P_\theta(x_i | x_{<i})}$ corresponds to the expected number of bits needed to encode each token. Accordingly, if each token was originally $b$ bits uncompressed, then the induced compression rate is $b / b_{\theta}$. Therefore, we can 
estimate compression performance directly from model loss without implementing the full arithmetic encoder.

\subsection{Standard LM Compression}

Here we formalize our baseline AR setup for audio waveforms. 
Each sample of uncompressed audio is a signed integer of $b$ bits from 
${\mathbb{Z}_{b} = \{z \in \mathbb{Z} \mid -(2^{b-1}) \leq z < 2^{b-1}\}}$, 
and a waveform $\mathbf{w} \in \mathbb{Z}_{b}^{T f_s \times c}$ is an array of samples 
where $T$ is duration in seconds, $f_{s}$ is sample rate, $c$ is number of channels, and $b$ is bit depth. 
We first convert signed samples to unsigned via ${\mathbf{x} = \mathbf{w} + 2^{b-1}}$, yielding ${\mathbf{x} \in \mathbb{N}_{b}^{T f_s \times c}}$, where 
${\mathbb{N}_b = \{z \in \mathbb{N} \mid 0 \leq z < 2^b\}}$.
We feed $\mathbf{x}$ through a standard decoder-only Transformer \cite{vaswani2017attention} with causal masking, similar to GPT-2 \cite{radford2019language}.
The model outputs probability distributions over the next sample $P_\theta(x_{i} \mid x_{< i})$ for all vocabulary tokens, and the training objective is to maximize the likelihood $\mathcal{L}_\theta = \sum_{i = 1}^{N} \log P_\theta(x_{i} \mid x_{< i})$, equivalent to minimizing cross-entropy loss.

In past work, each audio sample is treated as a token, inducing a vocabulary size is ${|\mathcal{V}| = 2^{b}}$ tokens.
While this works reasonably for 8-bit audio (${|\mathcal{V}| = 256}$), it becomes prohibitive at higher bit depths: 16-bit requires 65,536 tokens and 24-bit requires 16,777,216 tokens.
The embedding and output layers scale as $O(d \cdot 2^{b})$ parameters where $d$ is model dimension, quickly exceeding the size of the entire transformer backbone and creating intractable memory requirements.
The context window is limited by the Transformer's maximum sequence length (e.g., 2,048--8,192 samples, $\sim$50--200ms at 44.1kHz), necessitating sliding windows or chunking for longer audio.
This intractability at higher bit depths motivates the byte-level tokenization approaches we explore next.

\subsection{Trilobyte: Hierarchical Byte-Level Tokenization}

To address the vocabulary explosion of sample-level tokenization
we introduce Trilobyte, which decomposes each $b$-bit sample into $B = \ceil{b/8}$ bytes.
Rather than modeling these bytes with distinct subvocabularies, Trilobyte simply predicts over 256 possible values at each byte position in the sequence, maintaining a constant vocabulary size regardless of bit depth.
This reduces vocabulary scaling from exponential $O(2^{b})$ to constant $O(1)$ while enabling the model to implicitly learn separate distributions for each byte position autoregressively through the one-to-one mapping between byte position and sequence index.
We interleave the constituent bytes of each audio sample (MSB, middle byte(s), LSB), and train a GPT-2 architecture \cite{radford2019language} on these byte-level sequences.
We also experimented with an expanded vocabulary variant that partitions tokens into explicit subvocabularies per byte position (linear rather than constant scaling with bit depth), but this yielded negligible compression gains ($< 0.003$x), suggesting the constant vocabulary already learns separate byte distributions implicitly through AR context.

For stereo audio, we concatenate left and right channels in random order (either $x_1^L, x_2^L, \ldots, x_{N/2}^L, x_1^R, x_2^R, \ldots, x_{N/2}^R$ or vice versa) rather than interleaving at the sample level ($x_1^L, x_1^R, x_2^L, x_2^R, \ldots$), allowing the model to exploit cross-channel correlations when transitioning between channels in its AR predictions.
We found that compression rates are nearly identical between concatenation and sample-level interleaving, so we use concatenation for simplicity.
By conditioning the second channel's predictions on the first channel's context, the model can potentially capture redundancies beyond FLAC's mid-side encoding, which is limited by small block sizes.

Trilobyte enables the first tractable language model compression of 24-bit professional audio while achieving comparable performance to naive 16-bit modeling with full softmax.
Since Trilobyte operates on byte-level sequences, we compute bits per byte (BPB) and convert to compression rate as $8 / \text{BPB}$.
Note that at 8-bit, Trilobyte's tokenization collapses to standard sample-level tokenization (1 byte per sample, vocabulary of 256), so Trilobyte is identical to standard LMs for the 8-bit regime.
We compare Trilobyte against both sample-level tokenization and a byte-level in-context LM baseline.

\subsection{Additional Baselines and Experimental Approaches}

Beyond our primary ML-based approaches, we evaluate several additional baselines and experimental methods. 
We conduct extensive experiments with FLAC at different compression levels (0--8) to understand the performance ceiling of traditional methods across our diverse audio domains (see Appendix~\ref{app:flac} for detailed results).
As an additional baseline, we compare to the in-context compression approach of Delétang et al. \cite{deletang2023language} and Li et al. \cite{li2025lossless}, using the pretrained Llama-2-7B model \cite{touvron2023llama2}. 
Here, instead of training on audio, audio byte streams are compressed by naively encoding the bytes as gibberish text and compressing the text using the pre-trained LM.
As this method is intractably slow, following previous work \cite{deletang2023language} we report results over 1K randomly sampled, 1,024-sample chunks across each dataset. We discuss our further experiments with this approach in Appendix~\ref{app:lmic}.
Additionally, we explored neural audio codecs 
as drop-in replacements for linear predictive coding in FLAC-style compression, hypothesizing that learned representations might yield better residual distributions for Rice coding (Appendix~\ref{app:nac}).
While most of these approaches yielded poor results relative to our primary methods, they provide 
insights into the limits of different compression paradigms.

\section{Experimental Setup}\label{sec:experiments}

\begin{table*}[t]
    \centering
    \caption{
    Compression rates across methods and bit depths.
    FLAC uses compression level 8. 
    In-context uses pretrained Llama-2-7B \cite{touvron2023llama2}. 
    Standard refers to sample-level tokenization; Trilobyte uses hierarchical byte-level tokenization (both with trained Transformers).
    Sample-level is equivalent to Trilobyte at 8-bit and intractable at 24-bit (16.7M vocabulary).
    Transfer denotes a single 24-bit Trilobyte model trained on all datasets with lower bit masking for lower bit depths.
    Bold indicates best performance.
    }
    \newcommand{\resultrow}[9]{#8 & #1 & #2 & #7 & #5 & #4 & #6& #9\\}
    \begin{tabular}{ccll|cc|cc||c}
        \toprule
        \textbf{$b$ (Bits)} & \resultrow{\textbf{Domain}}{\textbf{Dataset}}{\textbf{FLAC (default)}}{\textbf{Standard}}{\textbf{In-context}}{\textbf{Trilobyte}}{\textbf{FLAC}}{\textbf{$f_{s}$ (Hz)} / \textbf{$c$ (\#Ch)}}{\textbf{Transfer}}
        \midrule
        \multirow{3}{*}{8}
            & \resultrow{Speech}{SC09}{0.94}{{2.08}}{1.80}{{2.08}}{0.95}{16000 / 1}{\textbf{2.88}}
            & \resultrow{Music}{Beethoven}{1.66}{\textbf{7.94}}{1.33}{\textbf{7.94}}{1.69}{16000 / 1}{7.45}
            & \resultrow{}{YouTube Mix}{1.56}{{4.15}}{1.28}{{4.15}}{1.58}{16000 / 1}{\textbf{5.14}}
        \midrule
        \multirow{8}{*}{16}
            & \resultrow{Speech}{VCTK}{2.29}{{2.66}}{1.75}{{2.66}}{2.32}{48000 / 1}{\textbf{2.68}}
            & \resultrow{}{LJSpeech}{1.65}{1.98}{1.49}{\textbf{2.08}}{1.69}{22050 / 1}{2.04}
            & \resultrow{}{LibriSpeech}{1.71}{2.06}{1.64}{\textbf{2.11}}{1.74}{16000 / 1}{2.10}
            & \resultrow{Bioacoustics}{Birdvox}{2.32}{2.47}{1.75}{\textbf{2.48}}{2.33}{24000 / 1}{\textbf{2.48}}
            & \resultrow{SFX}{Epidemic Sound}{2.50}{3.00}{1.70}{\textbf{3.40}}{2.63}{48000 / 1}{3.10}
            & \resultrow{Music}{MusDB18 (All)}{2.09}{2.64}{1.53}{\textbf{2.82}}{2.15}{44100 / 1}{2.75}
            & \resultrow{}{MusDB18 (Mixes)}{1.80}{1.85}{1.27}{\textbf{2.08}}{1.87}{44100 / 2}{1.98}
            & \resultrow{}{Commercial 16-bit}{1.72}{1.64}{1.26}{\textbf{1.86}}{1.74}{44100 / 2}{1.74}
        \midrule
        \multirow{1}{*}{24}
            & \resultrow{Music}{Commercial 24-bit}{1.61}{\textcolor{red}{\xmark}}{1.07}{1.48}{\textbf{1.63}}{44100 / 2}{{1.47}}
        \bottomrule
    \end{tabular}
    \label{tab:results}
\end{table*}

Our evaluation spans multiple datasets across diverse audio domains. 
For \textbf{music}, we use MusDB18 \cite{musdb18}, a multi-track music database with stereo mixes and stems in various configurations (stereo mixes, mono mixes, stereo stems, mono stems, combined stereo, combined mono). 
To better reflect real-world compression scenarios, we additionally include a dataset of 
commercial music data at multiple bit depths and quality levels (16-bit and 24-bit): 1,569 songs (120 hours) at 16-bit, and 933 songs (70 hours) at 24-bit, with the latter including high-resolution recordings up to 192kHz sampling rate. 
Unlike many academic music datasets, which are often distributed in processed or lossy formats and may not reflect the distribution of commercially released recordings, this music more closely represents the types of lossless audio files that users seek to store and compress in practice.
However, for consistency, we resample all commercial music data to 44.1kHz, the most common sample rate in the corpus.
We additionally incorporate music datasets Beethoven \cite{goel2022s,mehri2016samplernn} (recordings of Ludwig van Beethoven's piano sonatas) and YouTube Mix \cite{goel2022s,deepsound} (piano music from YouTube) into our experiments. 
For \textbf{speech}, we evaluate on LibriSpeech \cite{panayotov2015librispeech} (clean read English speech), LJSpeech \cite{ljspeech} (single-speaker audiobook recordings), SC09 \cite{goel2022s,donahue2019adversarial,Warden2018SpeechCA} (spoken digit recognition), and VCTK \cite{yamagishi2019cstr} (multi-speaker English corpus). 
We also evaluate on other audio domains: Birdvox \cite{farnsworth2022birdvox} (\textbf{bioacoustic} bird vocalizations) and Epidemic Sound \cite{epidemic} (\textbf{sound effects} library).
Although some files may contain lossy-coded artifacts, we utilize YouTube Mix, Beethoven Piano Sonatas, and SC09 for alignment with prior AR-based compression research \cite{goel2022s}.
However, many of our datasets remain truly uncompressed, and compressing previously lossy audio represents a realistic downstream scenario.

We evaluate at each dataset's native bit depth: 8-bit (Beethoven, YouTube Mix, SC09, all 16kHz), 16-bit (LibriSpeech 16kHz, LJSpeech 22.05kHz, Birdvox 24kHz, MusDB18 44.1kHz, VCTK and Epidemic Sound 48kHz, and commercial 16-bit at 44.1kHz), and 24-bit (commercial 24-bit at 44.1kHz).
We compare FLAC (compression level 8, maximum), sample-level tokenization (90M params at 8-bit, 140M at 16-bit),
and Trilobyte (90M). All trained models are trained for a fixed number of 300K steps.

\subsection{Results}

Table~\ref{tab:results} presents compression rates across all methods and bit depths.
Throughout this section, percentages refer to relative gains (e.g., 2x to 3x is a 50\% improvement, yielding 50\% smaller files).
At 8-bit, the standard and proposed Trilobyte tokenization schemes are equivalent: both substantially outperform FLAC, achieving 370\%, 163\%, and 119\% improvements on Beethoven, YouTube Mix, and SC09, respectively.
The wide variation in Trilobyte's 8-bit performance (2.08--7.94x) indicates that compression gains depend heavily on audio domain structure---here, the music datasets are acoustically narrow (solo piano) and compress better than SC09 (multi-speaker, multi-microphone). 

At 16-bit, AR modeling provides consistent compression gains relative to FLAC, 
though the gains are modest compared to those seen for 8-bit audio:  
15\%, 31\%, and 21\% improvements on VCTK, MusDB18 Mono, and LibriSpeech, respectively.
Moreover, at 16-bit, FLAC compression rate correlates to Trilobyte compression rate across datasets ($r = 0.92$, $p \ll 0.01$). 
Higher sample rates appear to have less of an effect on overall compression rate relative to bit depth---some high sample rate datasets compress better than low rate ones (VCTK at 48kHz is 2.66x vs. LibriSpeech at 16kHz is 2.11x). 
Sample-level tokenization performs comparably to Trilobyte on some datasets (2.66x on VCTK, 2.47x on Birdvox, 2.64x on MusDB18 Mono) but generally falls short, particularly on music.
Notably, Epidemic Sound achieves 3.40x compression with Trilobyte, a 29\% improvement over FLAC.

At 24-bit, bit depth becomes the fundamental barrier: the 16.7M-token vocabulary required for sample-level tokenization is completely intractable, requiring approximately 12B parameters for the output projection matrix alone.
Trilobyte's hierarchical byte-level decomposition sidesteps this exponential scaling by reducing the vocabulary to just 
256 
tokens, enabling the first tractable 24-bit language model compression. 
However, 
our approach (1.48x) falls 9\% short of FLAC (1.63x). 
One possible explanation is that a non-trivial amount of the information in the least significant bits at $24$-bit is imperceptible noise---audio tool chains with up to $144$dB of dynamic range are needed to preserve the signal at $24$ bits. 
Rice coding in FLAC may be nearly optimal for compressing this low amplitude noise.
The in-context LM approach with pretrained Llama-2-7B underperforms both FLAC and Trilobyte across all datasets and bit depths except for 8-bit SC09, suggesting that text-pretrained models without domain-specific training struggle to capture audio structure effectively.

\subsubsection{Transfer Learning with Trilobyte}

We further investigated Trilobyte's ability to losslessly compress \emph{arbitrary bit depths} with a \emph{single} model. 
Specifically, by masking out the lower bytes with a learned null token, we can simultaneously model multiple bit depths within a single model during training, synthetically interleaving lower bit-rate audio with such null tokens at inference to perform any-bit lossless compression.
To evaluate this capability, we train models on Commercial (24-bit) and MusDB18 Stereo (16-bit) with an additional mask token and randomly drop out lower-significance bytes during training ($p = 0.1$), then test whether the resulting models can compress audio at multiple bit depths without retraining. 
For 24-bit Commercial music, we attain 1.49x (24-bit), 1.78x (16-bit), and 3.4x (8-bit) compression rates with a single model, while on 16-bit MusDB18 stereo we attain 2.07x (16-bit) and 3.8x (8-bit) compression. 

From this initial result, we then trained a \emph{single} Trilobyte model over all data using our multi-bit-rate masking scheme, with results shown in the final column of Table~\ref{tab:results} (\textbf{Transfer}). 
We find that the compression performance is similar to the per-dataset Trilobyte models (with our joint model performing slightly worse on some datasets and better on others), showing that we can successfully train a single generalist LM-based compressor (even without scaling the model size) over diverse audio corpora using Trilobyte and our lower byte masking.
We release this generalist Trilobyte model as an open-source lossless audio codec\footnote{\texttt{\url{https://github.com/pnlong/trilobyte-lossless-codec}}} to serve as a baseline for future research in learned lossless compression across diverse audio domains, sample rates, and bit depths.

\section{Conclusion}\label{sec:conclusion}

Our evaluation reveals that Trilobyte achieves consistent, albeit modest, gains over FLAC at 16-bit, with an average improvement of 18\% across domains. 
These modest 16-bit gains contrast sharply with 8-bit performance (217\% average improvement).
At 24-bit, Trilobyte trails FLAC by 9\% but enables the first tractable LM compression at this bit depth where sample-level approaches are completely infeasible; we hope to close this gap in future work.
Transfer results show a single 24-bit Trilobyte model trained on all datasets achieves compression rates comparable to dataset-specific models across bit depths.
Our results also highlight that bit depth, not sampling rate or data domain, becomes the primary bottleneck, as LMs consistently outperform FLAC at 8-bit but gains quickly diminish at higher bit depths. 
This suggests FLAC operates near fundamental entropy bounds for full-fidelity audio and our empirical compression rates establish lower bounds across audio domains.

We acknowledge that our ML approaches are orders of magnitude slower than FLAC---their modest compression wins are unlikely to justify their computational costs for real-world deployment.  
Nevertheless, this work addresses a critical gap in the literature: prior LM-based compression research has been constrained to 8-bit audio, leaving unexplored whether these methods scale to the full-fidelity regimes where lossless compression is actually needed.
We provide the first comprehensive benchmark of language model compression on CD-quality (16-bit) and professional (24-bit) audio, demonstrating that standard sample-level approaches face increasingly intractable vocabulary.
Trilobyte's hierarchical byte-level tokenization overcomes this fundamental barrier, reducing vocabulary scaling from exponential $O(2^{b})$ to 
constant $O(1)$ 
and enabling tractable modeling at arbitrary bit depths.
Although current compression gains remain modest, our work demonstrates that learned approaches can consistently outperform FLAC across diverse audio domains and bit depths. We anticipate that future research may work to scale the performance of such models and/or improve their efficiency.

\ifcameraready
\section{Acknowledgments}

We are grateful to Roger Dannenberg, Sander Dieleman, Jesse Engel, John Thickstun, Albert Gu, and Isaac Liao for helpful conversations about this work, which helped shape the direction of our research.
\fi

\section{Generative AI Use Disclosure}

Generative AI tools (Claude) were used solely for minor editing and formatting assistance.
All technical content, experimental work, and analysis were performed entirely by the authors, who are fully responsible for the work presented.

\bibliographystyle{IEEEtran}
\bibliography{references}

\appendix

\section{Extended FLAC Experiments}\label{app:flac}

\begin{figure}[!b]
    \centering
    \includegraphics[width=1.0\columnwidth]{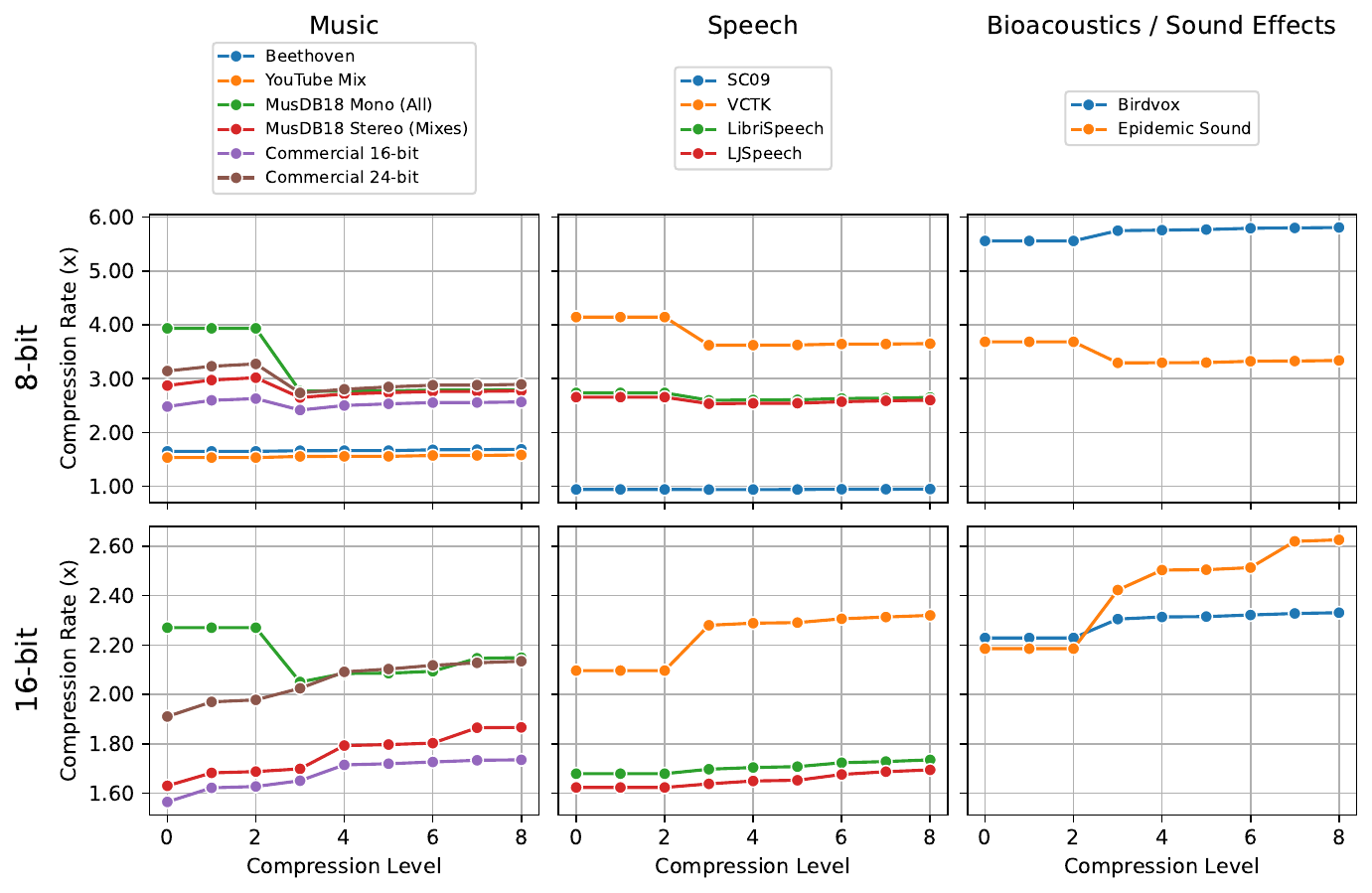}
    \caption{
        FLAC compression performance across diverse audio domains at 8-bit and 16-bit quantization levels. 
        Birdvox achieves exceptional compression ($\sim$6x at 8-bit), perhaps reflecting the sparse and structurally constrained nature of bird vocalizations, which are highly predictable under linear predictive coding. 
        Meanwhile, speech and music datasets show more modest gains. 
        16-bit audio generally achieves 1.5--2.5x compression, with diminishing returns beyond FLAC level 3.
        Note that we disable FLAC's verbatim, constant, and fixed subframe types, and that we do not evaluate Beethoven, YouTube Mix, or SC09 beyond 8-bit because they are 8-bit datasets.
    }
    \label{fig:flac}
\end{figure}

FLAC offers compression levels 0--8, where higher levels exhaustively try more subframe types, linear predictive coding orders, and Rice parameters to find optimal compression at the cost of slower encoding (decoding speed remains constant). 
Level 5 (the default) provides a good balance between compression quality and speed, while level 8 tries essentially all combinations. 
Figure~\ref{fig:flac} shows FLAC compression rates at different compression levels (0--8) for 8-bit and 16-bit audio across all datasets. 
An important configuration note: to isolate the effect of linear predictive coding as a lossy estimator, we disable FLAC's verbatim, constant, and fixed subframe types, since one could trivially implement these same subframe types in any audio codec.
This explains the noticeable drop in compression rate at level 3, especially visible in MusDB18 Mono, over 80\% of which consists of sparse, largely-silent stems.
These subframe types are normally used to efficiently encode silence and constant-value regions; without them, silent regions compress less effectively.

Compression rate patterns vary significantly by bit depth. 
For 8-bit audio, compression varies widely by dataset. 
As a whole, music achieves quite moderate compression (1.8--3x), though MusDB18 Mono in particular achieves 4x due to sparse multi-track content. 
Speech datasets LibriSpeech, LJSpeech, and VCTK achieve 2.5--4x compression while SC09 seems to not compress, perhaps a consequence of the short durations of tracks in the dataset. 
Remarkably, Birdvox achieves exceptional $\sim$6x compression, likely due to sparse yet locally-predictable bioacoustic signals. 
For 16-bit audio, compression is more modest, generally achieving 1.5--2.7x compression. 
Music datasets cluster around 1.6--2.3x compression, speech datasets achieve roughly the same, while bioacoustics and sound effects perform the best, with Epidemic Sound reaching 2.6x compression.

\section{Neural Audio Codecs for Compression}\label{app:nac}

\begin{figure}[t]
    \centering
    \includegraphics[width=1.0\columnwidth]{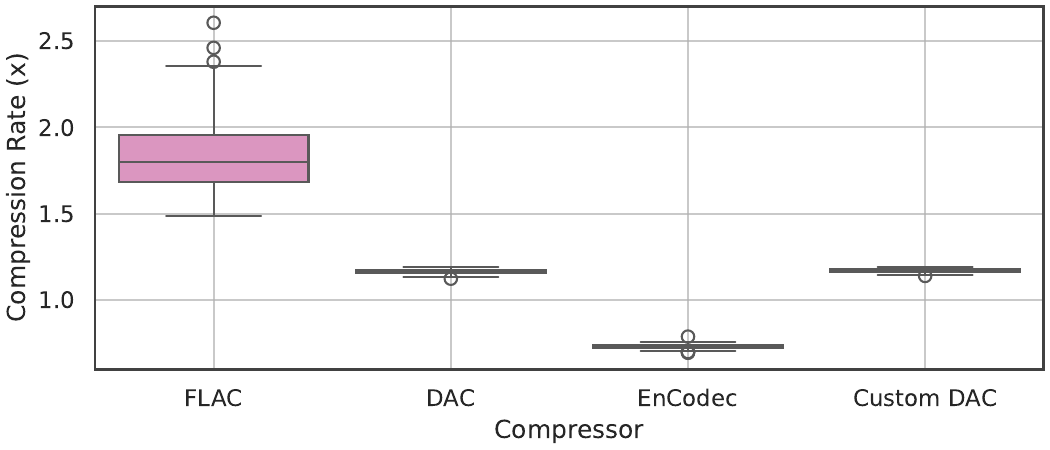}
    \caption{
        Compression rate comparison across FLAC, DAC, EnCodec, and Custom DAC compressors on MusDB18 mixes. 
        FLAC achieves the best compression, at approximately 1.8x, while the NAC-based approaches underperform, with EnCodec actually increasing file size.
    }
    \label{fig:lnac}
\end{figure}

We also tried the alternative paradigm of FLAC-style compression: a bottleneck representation plus residuals encoded with Rice coding \cite{rice1979some}. We explored neural audio codecs (NACs) as drop-in replacements for linear predictive coding in this pipeline. This approach did not outperform FLAC; we summarize the setup and results here.

FLAC compresses audio using linear predictive coding and Rice-coded residuals; we replace linear predictive coding with a NAC---Descript Audio Codec (DAC) \cite{kumar2023high}, EnCodec \cite{defossez2022high}, or Custom DAC---storing latent codes and Rice-encoded residuals. The compression pipeline is to (1) encode audio through the NAC encoder to obtain discrete latent codes $\mathbf{z}$, (2) decode to obtain reconstruction $\hat{\mathbf{x}}$, (3) compute residuals $\mathbf{r} = \mathbf{x} - \hat{\mathbf{x}}$, and (4) store $\mathbf{z}$ and Rice-encoded $\tilde{\mathbf{r}}$. Decompression reverses this process. Rice coding assumes residuals follow a roughly geometric distribution.

We evaluate on MusDB18 \cite{musdb18} mixes (44.1kHz, 16-bit CD-quality stereo music). We compare four approaches: \textbf{FLAC} (at the default compression level 5), \textbf{DAC} (DAC-44.1kHz with 3 codebook levels), \textbf{EnCodec} (EnCodec-48kHz with 4 codebook levels), and \textbf{Custom DAC} (our DAC-44.1kHz variant trained without adversarial or perceptual losses). We hypothesize that our Custom DAC variant will produce residuals more amenable to Rice coding than NACs trained purely for perceptual quality.
Figure~\ref{fig:lnac} shows compression rates. FLAC achieves the best compression (approximately 1.8x). DAC and Custom DAC underperform at about 1.2x; EnCodec performs worst, actually \emph{increasing} file size (compression rate $<$ 1.0x). Several factors explain NAC underperformance, including that NAC residuals do not follow the geometric distribution Rice coding assumes.

\begin{figure}
    \centering
    \includegraphics[width=1.0\columnwidth]{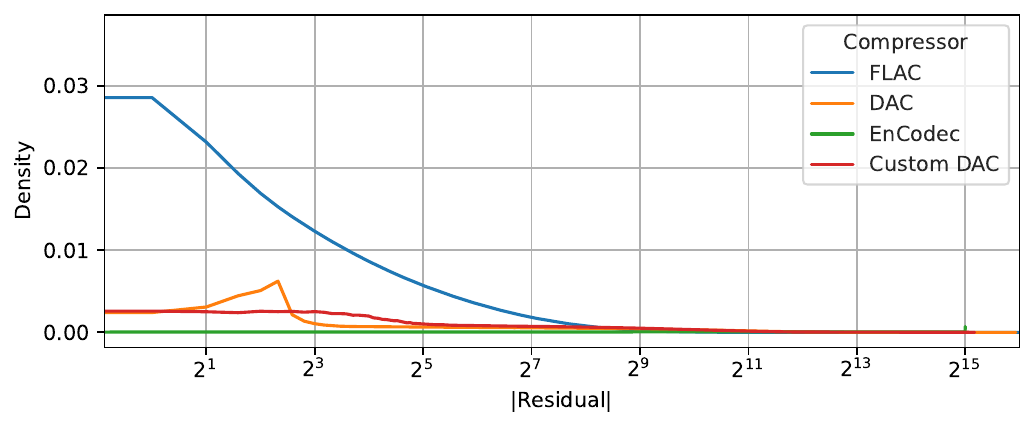}
    \caption{
        Residual distribution comparison showing residual magnitudes (note the log scale) for FLAC, DAC, EnCodec, and Custom DAC compressors. 
        FLAC residuals follow a geometric distribution 
        with a mean absolute residual of 156.34, while DAC, EnCodec, and Custom DAC residuals are more uniformly distributed regardless of codebook level, with mean absolute residuals of 1,603.54 (DAC), 18,376.66 (EnCodec), and 1,245.76 (Custom DAC) -- an order of magnitude larger than FLAC.
    }
    \label{fig:residuals}
\end{figure}

Figure~\ref{fig:residuals} shows histograms of absolute residuals for FLAC, DAC, EnCodec, and Custom DAC compressors. 
FLAC residuals exhibit a clear geometric distribution (concentrated near zero with exponentially decreasing frequency), precisely what Rice coding requires. 
In contrast, NAC-based residuals are more uniformly distributed with less concentration near zero, deviating from the geometric assumption.

This distributional mismatch is critical. 
Rice coding's efficiency relies on the geometric distribution assumption: when residuals follow different distributions (as with NAC reconstructions), Rice coding becomes suboptimal. 
This fundamental incompatibility explains why NAC-based approaches do not outperform FLAC despite potentially capturing a more complex audio structure. 
The underlying cause is that neural codecs are trained with adversarial perceptual losses that optimize for human-perceivable quality rather than minimizing absolute waveform error, creating structured reconstruction errors that do not follow simple geometric distributions. 
Additionally, EnCodec's extremely poor compression performance is explained: its distribution of residuals appears uniform, and certainly not geometric.

Notably, our training of a custom DAC-44.1kHz variant without adversarial or perceptual losses (Custom DAC) appears to have produced a slightly different residual distribution from DAC itself, as our Custom DAC compressor has a markedly lower mean absolute residual. 
However, this improvement does not translate into better compression rates, likely because the Custom DAC's residual distribution remains non-geometric. 

NAC-based approaches fail to outperform FLAC on both compression rate and speed. The fundamental issue is that NAC residuals do not follow geometric distributions, which makes Rice coding inefficient.

\section{In-context LMs for Compression}\label{app:lmic}

\begin{figure}
    \centering
    \includegraphics[width=1.0\columnwidth]{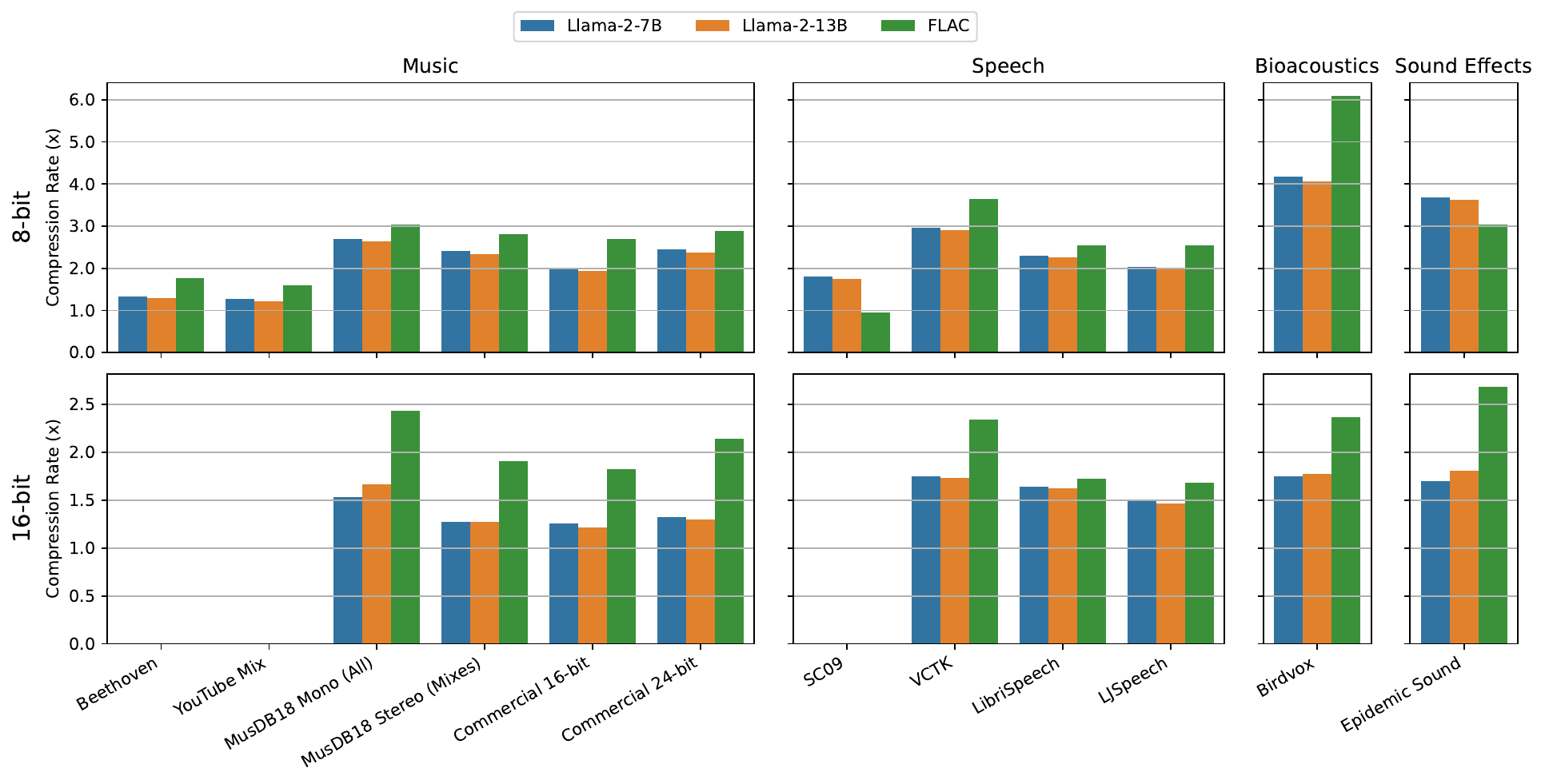}
    \caption{
        In-context LM-based compression performance with the method defined in Delétang et al. \cite{deletang2023language} and Li et al. \cite{li2025lossless} using pre-trained language models (Llama-2-7B and Llama-2-13B \cite{touvron2023llama2}) across diverse audio domains at 8-bit and 16-bit quantization. 
        We also report FLAC compression results at compression level 8, the maximum. 
        Model scaling (7B to 13B) shows minimal gains at 8-bit and some improvements at 16-bit, especially for complex datasets. 
        This method underperforms FLAC on most signals, with the exception of SC09 and Epidemic Sound at 8-bit.
    }
    \label{fig:lmic}
\end{figure}

Figure~\ref{fig:lmic} shows compression rates for the in-context LM-based approach from Delétang et al. \cite{deletang2023language} and Li et al. \cite{li2025lossless} using Llama-2-7B and Llama-2-13B \cite{touvron2023llama2} across all datasets for 8-bit and 16-bit audio. 
Because this method of compression is intractably slow, the results shown here are over 1,000 randomly-selected, 1,024-sample chunks (1,024 bytes for 8-bit, 2,048 bytes for 16-bit) across each dataset. 
Model scaling effects differ by bit depth. 

Comparing the language model-based approach to FLAC reveals that it underperforms in nearly all scenarios. 
FLAC dominates across most datasets and bit depths: for Birdvox at 8-bit, FLAC achieves $\sim$6x compression compared to $\sim$4x for the in-context LM approach, and for music datasets like MusDB18 Mono and the commercial data, FLAC consistently outperforms by substantial margins. 
The only exceptions are SC09 at 8-bit, where the language model approach ($\sim$1.8x) significantly outperforms FLAC (1x, effectively no compression), and Epidemic Sound at 8-bit, where the language model approach achieves $\sim$3.5x compared to FLAC's $\sim$3x. 
Model scaling from 7B to 13B parameters shows minimal gains, and in fact, often leads to lower compression rates.

The broader implications highlight fundamental limitations of this approach for lossless audio compression. 
As Delétang et al. \cite{deletang2023language} acknowledge, the model size (in bytes) must be amortized over the compressed data, which drastically lowers the effective compression rates---a 7B parameter model requires gigabytes of storage, making it impractical unless compressing massive audio archives. 
The computational cost of running inference with billion-parameter models is only justified in domains where they significantly outperform FLAC, which our results show are incredibly limited. 
For the vast majority of audio domains including music and bioacoustic signals, FLAC remains the superior choice, suggesting that general-purpose language models trained on text do not transfer effectively to the diverse statistical structures present in audio waveforms.

\end{document}